# A Logical Formalization of a Secure XML Database[1]

Alban Gabillon

Université de Pau et des Pays de l'Adour, IUT de Mont de Marsan, LIUPPA/CSySEC,
40 000 Mont de Marsan, France.
alban.gabillon@univ-pau.fr

**Abstract**. In this paper, we first define a logical theory representing an XML database supporting XPath as query language and XUpdate as modification language. We then extend our theory with predicates allowing us to specify the security policy protecting the database. The security policy includes rules addressing the read and write privileges. We propose axioms to derive the database view each user is permitted to see. We also propose axioms to derive the new database content after an update.

## 1 Introduction

Several discretionary access control models for eXtensible Markup Language (XML) documents [2] have been proposed [1][7][13][11][14]. In [10], we first reviewed most of the existing access control models for XML and discussed their weaknesses. Based on our study, we then defined a new model suggesting a solution to cope with the problems that our study revealed. The model in [10] has the following characteristics:
- The model is an interpretation for XML of the SQL security model (with some additional features).
- Each user is provided with a view of the source database he or she is permitted to see.
- The model includes a new position privilege that allows knowing about the existence of an XML node but not about its label. Nodes on which users hold a position privilege are shown with a RESTRICTED label in users' views. Thus, sensitive labels are hidden while the structure of the XML document is preserved.
- The model includes various write privileges and defines the access controls for the write operations.

However, the model in [10] has two drawbacks:
- It is defined in an informal way.
- Like the security model of SQL, it ignores interactions between the read privilege and the write privilege. In other words, write operations are evaluated on the source database and not on the user's view. As a consequence, users can build some covert channels to learn about the data they are not permitted to see.

---

[1] This work was supported by funding from the French ministry for research under "ACI Sécurité Informatique 2003-2006. Projet CASC".

In [28], we updated the model defined in [10] and we obtained a new version of our model which did not have these drawbacks:
- We investigated in details interactions between the read privilege and the write privilege.
- We used mathematical logic to formally define the model. With logic, we could homogeneously define the database, the security policy and the access controls. This approach enables an easy and precise way of computing the facts that users are authorized to see/update.

However, due to space limitations, the logical formalization in [28] was not complete:
- We could not give the logical interpretation of the query language.
- We could not give the axioms allowing us to derive the various tree geometry relations (child, descendant, following-sibling …).
- We could not describe the numbering scheme we used to assign identifiers to XML nodes.

Therefore, the purpose of this paper is to give a full logical definition of an XML database secured with the model we first defined in [10] and later refined in [28]. In fact, this paper is an extended version of papers [10] and [28].

We have implemented a Prolog-based prototype simulating a secure XML database. Logical formulae given in this paper are Horn clauses and have been implemented as such in the prototype. This prototype can be downloaded from the following address: http://www.univ-pau.fr/~gabillon/xmlsecu

In section 2, we make an informal overview of our model and we underline the main limitations of existing security models for XML. In section 3, we define the logical theory representing an XML database. Since we use XPath [4] as a query language and XUpdate [15] as a modification language, we give their logical interpretation. In section 4, we extend our theory with predicates allowing us to define the security policy protecting the database. The security policy includes rules addressing read and write privileges. We define the logical formulae allowing us to derive the database view each user is permitted to see. We also give the logical formulae allowing us to derive the new database after an update. Finally, section 5 concludes this paper.

## 2  Informal Overview of our Model

In this section, we survey existing works and we present our approach in an informal way. The interested reader can also refer to [9] for a formal presentation of existing works.

### 2.1  Subjects

The development of an access control system requires the definition of the subjects and objects. There is not much to say about the subjects in existing access control models for XML data. In [1], subjects are simply users. In [13], they can be users or roles (see [17] for a description of roles). In [11], they are users who are structured into a group hierarchy. In [7], the authors take into consideration the fact that their access control model is to be implemented as an extension to an existing web server. In their model, subjects are users or `IP` addresses (or a combination of the two), and they are structured into a group hierarchy. In the model presented in this paper, subjects can be users or roles like in the `SQL` security model.

## 2.2 Objects

An object is a granule of information which can be protected. Regarding the objects, there are basically different approaches among existing access control models for `XML` data:
- In [7][13], the smallest object is an element. Authorizations specified for an element are intended to be applicable to the element itself, its content (`PCDATA`) and its attributes.
- In [1], the smallest object is an element or an attribute. Authorizations specified for an element are intended to be applicable to the element itself and its content.
- In [11], an object is a node of an `XPath` tree (see [4] for a description of `XPath` and figure 1 for an example of an `XPath` tree). A node of an `XPath` tree can be an element, the content of an element or an attribute.

We want our security model to have a high expressive power. Therefore, in our model, an object is a node of an `XPath` tree like in the model in [11].

## 2.3 Security Policy

In all the existing models, the approach is the same: the security administrator writes the security policy in a separate authorization sheet. The security policy consists of a set of authorization rules which can be either positive (grant) or negative (deny). The reason for having both positive and negative authorizations is to have a way to specify exceptions to authorizations which are applicable to sets of subjects or objects. Authorization rules address granules by means of `Xpath` expressions. Conflicts between the rules are solved by a conflict resolution policy.
- The model in [7] supports the `read` privilege only. In another paper [29], the authors partially address the `write` privilege but underline the fact that current `XML` applications are mostly read-only and that no consensus has emerged up to now on a model for `XML` updates. An authorization specified on an element can be defined as local. In that case, it is applicable to the element itself, its content and its attributes. It can also be defined as recursive. In that case, the permission/prohibition is propagated to the sub-elements. Authorizations can be specified at the document-level or at the Document Type Definitions (`DTDs`) level. Authorizations specified at the `DTD`-level propagate to all `XML` documents that are instances of that `DTD`. The conflict resolution policy applies "the most specific object takes precedence" principle. According to this principle instance-level authorizations override `DTD`-level authorizations and a recursive authorization propagates until overridden by a conflicting authorization on a more specific object. For conflicts which cannot be solved by this principle, the authors suggest to apply other principles like "the most specific subject takes precedence" or "denial takes precedence" …etc.
- The model in [1] supports two kinds of privileges: `browsing` (`read`) and `authoring` (`write`). Authorizations are specified along with propagation options. Depending on its propagation option, an authorization referring to an element may propagate to all the direct and indirect sub-elements, propagate to all the direct sub-elements only, or not propagate at all. Like in the previous model, authorizations can be specified at the `DTD`-level or at the instance-level. The conflict resolution policy is similar to the previous one.
- The model in [13] supports `read` and `write` privileges. The authors define three types of propagation policies: no propagation, propagation up (an authorization referring to an element is propagated to all its parent elements) or propagation down (an authorization referring to an element is propagated to all its sub-elements). The conflict resolution policy is either "denials take precedence" or "permissions take precedence". The main contribution of this paper is to propose a provisional authorization model. A provisional authorization is more than a simple permission/denial. Typically, a provisional authorization specifies that a user has to perform a given action (obligation) before he/she is granted a given privilege
- The model in [11] supports the `read` privilege only. The authors do not define any propagation policy. The conflict resolution policy is based on the priority of the different rules.
- The model in [14] is based on the model defined in [7]. The authors add `write` privileges and suggest a technique for efficiently managing access controls in a web environment which emphasizes the integrity of the documents (i.e. validity with respect to `DTDs`). They consider `XQuery` update operators but these operators are not standardized yet [20][3].

The model presented in this paper supports `read` and `write` privileges. We assume that security rules are issued in the chronological order. In case of a conflict between two rules, the last issued rule has the highest priority. The default policy is always "denial".

### 2.4 View Access Control

Most of the existing security models for `XML` define a view-based access control strategy for handling the `read` privilege. However, these models suffer from problems that were pointed out in [18]:
- Regarding the model in [11], if access to a node is denied then the user is not allowed to access the entire sub-tree under that node even if access to part of the sub-tree is permitted, therefore limiting the availability of data.
- Regarding the model in [7], in order to preserve the structure of the document, the authors allow elements with negative authorizations (i.e. access denied) to be released if the element has a descendant with a positive authorization (access permitted), thus making the semantics of the negative authorization unclear.

These two problems are illustrated in figures 1 and 2. Let us assume tag + represents permission and tag – represents denial for a given user `s`.

Figure 1 is related to the model in [11]. Recall that this model allows node-level security granularity. The left tree represents a sample medical files database which has been tagged according to the authorizations applying to a given user s. The right tree corresponds to the view user s is permitted to see. This view has been computed according to the access control strategy defined in [11]. One can see that the sub-tree of which element node /robert is the root is not visible at all although user s has the permission to see some of the descendant nodes. As it is said in [18], this access control strategy reduces the availability of data. One could argue that such a situation comes from a bad security design. Nevertheless, it should be possible to deny access to an internal node while granting access to its descendant nodes.

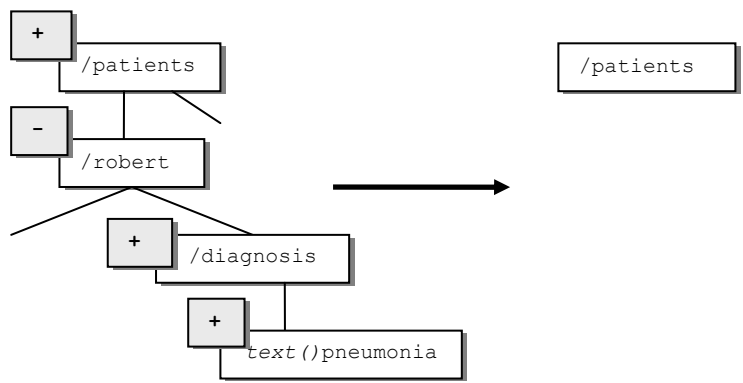

**Fig. 1. View Access Control in [11]**

Figure 2 is related to the model in [7]. Recall that this model allows element-level security granularity. The left tree represents the source tree which has been tagged according to the authorizations applying to user s. The right tree corresponds to the view user s is permitted to see. This view has been computed according to the access control strategy defined in [7]. One can see that element /robert is visible in the view. This makes the semantics of the negative authorization unclear.

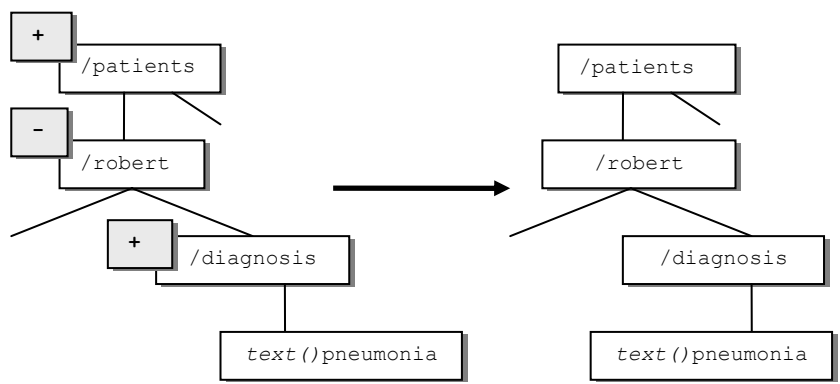

**Fig. 2. View Access Control in [7]**

In order to solve the problems mentioned in [18], we introduced in [10] a special privilege protecting the *existence* of nodes. Therefore, our model includes two kinds of read privileges: one privilege which allows knowing about the existence of a node (we call it the `position` privilege) and another privilege which allows knowing both the existence and the label (we call it simply the `read` privilege). The security designer has now two options:

- if user `s` is forbidden to know about the existence of node `n` then the security designer denies both `position` and `read` privileges on node `n` to user `s`. In that case node `n` and possible descendant nodes (even those for which the user has permission to read) are not shown in the view user `s` is permitted to see.
- if only the label of node `n` is sensitive then the security designer grants to user `s` the `position` privilege on node `n` (without granting the `read` privilege). Node `n` is shown in the view with RESTRICTED label and descendant nodes for which the user has permission to see are also shown in the view. Label RESTRICTED was first used by Sandhu and Jajodia in the context of multilevel databases [19]. Its semantics is "the label exists but you are not allowed to see it".

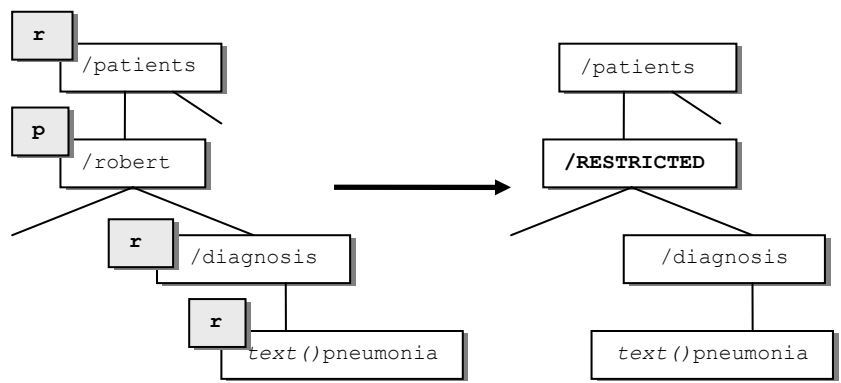

**Fig. 3. View Access Control in [10]**

In figure 3, tag **r** (respectively **p)** attached to a node represents the fact that user `s` holds the `read` (respectively `position`) privilege on that node. Right tree represents the view user `s` is permitted to see. User `s` is permitted to read illnesses (most probably for statistical purpose) but she is forbidden to see patients' names.

### 2.5 Write Access Controls

Some of the existing security models for XML consider the `write` privilege but,
- they do not clearly indicate in which framework the different update operations for XML are supported,
- the access control strategy that they use for handling the `write` privilege is not clearly described,
- interactions between read access controls and write access controls are not investigated.

In fact, these security models were designed to be implemented as extensions to existing web servers.

SQL ignores interactions between the read privilege and the write privilege. Indeed, if a user submits a write operation (via one of the standard SQL commands: INSERT, UPDATE or DELETE) then this operation is evaluated on the source database and not on the data the user is permitted to read. As a consequence, users can easily learn about the data they are not permitted to see. For example, consider user_A who is the owner of the employee table and who has granted to user_B the sole update privilege on employee.

user_B is not permitted to see user_A's employee table,

```
SQL> SELECT * FROM user_A.employee;
ERROR ORA-01031: insufficient privilege
```

but user_B is permitted to update user_A's employee table:

```
SQL> UPDATE user_A.employee SET salary=salary+100 WHERE salary > 3000;
2 rows updated
```

Although user_B is not permitted to see user_A's employee table, she has been able to learn, through an update command, that there are two employees with a salary greater than 3000. The UPDATE command was evaluated on data user_B was not permitted to see. Note, in particular, that the WHERE clause performed a read operation on the employee table. We could show various examples exploiting this vulnerability.

The model in [10] has the same vulnerability since it is an interpretation for XML of the SQL security model. In [10], an operation updating XML data is evaluated on the source database regardless of the read privileges held by the user submitting the operation.

In [28], we investigated interactions between the read privilege and the write privilege and we modified our model as follows: since a write operation is a process running on behalf of a user, it should have the privileges and the limitations of the user. In particular the write operation should not be able to read the data the user is not permitted to see. Conceptually, this means that write operations have to be evaluated on users' views and not on the source database. In order to retrieve the corresponding database node from a given view node, we assign one identifier to each database node. Identifiers are obtained by applying a numbering scheme [6][24][8][12][21][25][26]. If a user submits a write operation then the nodes to update are first selected from the view she is permitted to see. Thanks to the identifiers, corresponding database nodes are then retrieved and updated.

Our model supports three kinds of write privileges (insert, delete and update). We give the exact semantics of each of these privileges. We state the privileges that each XUpdate operation requires for completion and we formally define the access controls for each of the XUpdate operations.

## 3. XML Database

Mathematical logic has been used to formalize databases in two main directions. These directions are usually called the proof theoretic approach and the model theoretic approach. The former represents a database as a logical theory; the latter represents a database as an interpretation of a logical theory [16]. In this paper, we adopt the proof theoretic approach, that is, each database is associated with its logical theory `db`. We also make the *closed world assumption*. The closed world assumption holds that anything that we cannot show to be true is false.

### 3.1 XML documents modeled as trees

For the sake of simplicity, we shall not consider the type of XML nodes. An XML document is a *tree* of *nodes*. Each node is the parent of zero or more *child* nodes. Each node has one and only one parent except one fundamental node which has no parent and which is called the *document* node. We state that each node is associated with a *unique identifier* and a *label*:
− Node identifiers are obtained by applying a *numbering scheme* (see section 3.2).
− Labels are the data. Labels are small for nodes of type element (in the XML terminology they are referred to as *names*) and they can be very large for nodes of type text (in the XML terminology they are referred to as *values* or PCDATA).

Figure 4 shows an XML document which we shall use throughout the remainder of this paper. Strings `patients, franck, service, otolarynology` ... are labels. $n_1$, $n_2$, $n_3$, ... denote numbers identifying nodes. The document node has identifier `/` and label `/`. The document node has only one child node which is called the *root element* node. Identifier of the root element is $n_1$ and label is `patients`.

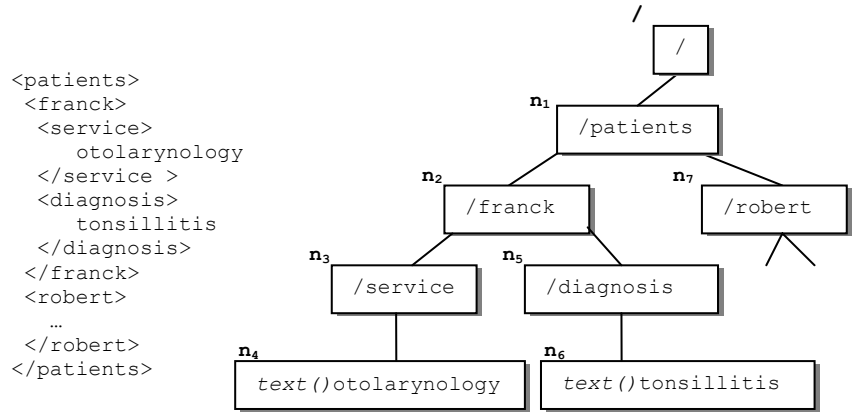

**Fig. 4.** XML document

### 3.2 Numbering Scheme

Several numbering schemes have been proposed [6][24][8][12][21][25][26]. They all support the representation of ancestor and sibling relationships between nodes i.e. one can derive the relationship between any two given nodes by looking at their unique numbers.

Some numbering schemes are *persistent* [12][8][25][26]. A persistent numbering scheme does not require renumbering after an update i.e. numbers assigned to existing nodes remain the same even after an update modifying the tree structure.

The scheme presented in this section is the scheme we defined in [12]. It is a very simple persistent numbering scheme which supports an infinite number of insertions without renumbering. It does not require a space of reserved identifiers like in [25]. It is based on the fact that there exist an infinite number of rational numbers within an interval `[a,b]`, `a`, `b` being rational numbers.

### 3.2.1 Static Numbering Step

We identify nodes by using rational numbers. We represent a rational number by a pair which consists of a signed integer and a strictly positive integer, e.g. we represent rational number `5/2` by the pair `(5,2)` and the rational number `-5/2` by the pair `(-5,2)`.

Our numbering scheme needs modest storage capacities. We assign a quintuple of integers `(l,(n_p,d_p),(n,d))` to each node:
- `l` is the level of the node in the tree.
- `(n,d)` is the local code of the node. Pair `(n,d)` represents the `n/d` rational number.
- `(n_p,d_p)` is the local label of the parent node.

*For a given level, local codes are unique.*

We assign the special identifier `(0,/,(1,1))` to the root node. `0` is the level. `/` is the code of the parent node (i.e. the document node). `(1,1)` is the local code.

Given a level `l`, if we assume that nodes are visited from left to right then the local code of a node is `(i,1)`, with `i` being the position of the node at level `l`.

After the static numbering step, node identifiers of the tree in figure 4 are the followings:

$$\begin{Bmatrix} n_1 = \bigl(0,/,(1,1)\bigr), n_2 = \bigl(1,(1,1),(1,1)\bigr), n_7 = \bigl(1,(1,1),(2,1)\bigr), n_3 = \bigl(2,(1,1),(1,1)\bigr), n_5 = \bigl(2,(1,1),(2,1)\bigr), \\ n_4 = \bigl(3,(1,1),(1,1)\bigr), n_6 = \bigl(3,(2,1),(2,1)\bigr) \end{Bmatrix}$$

### 3.2.2 Dynamic Numbering Step

In this section, we show how a newly inserted node is dynamically numbered without changing the number of existing nodes.

*Rules for creating identifiers for new nodes*
- If `v` is the first node to be inserted at level `l` then its local code is `(1,1)`
- If `v` is inserted immediately before the node of local code `(i,j)` and if there is no other node before `(i,j)` then the local code of `v` is `(i-j,j)`.
- If `v` is inserted immediately after the node of local code `(i,j)` and if there is no other node after `(i,j)` then the local code of `v` is `(i+j,j)`.
- If `v` is inserted immediately before the node of local code `(i,j)` and immediately after the node of local code `(k,h)` then the local code of `v` is `(a,b)` with `a=(i.h+k.j)\d` and `b=2.h.j\d`. `\` denotes the integer division. `d` is the highest common factor of `(i.h+k.j)` and `2.h.j`.

If, for example, node `/albert` is inserted between node `/franck` and node `/robert`, then its identifier will be `(1,(1,1),(3,2))`.

### 3.3 Language

Language L of theory db is based on first-order logic with equality. For the sake of simplifying our logical formulae, we shall consider that the database may contain only one document. We shall use the following two-place predicate to represent the database content:

− node(n,v) reads "there is a node with label v identified by number n"

We shall also use the following predicate to learn about the database tree geometry:

− child(x,y), reads "node[2] x is a child of node y"

There are also other tree geometry predicates like parent, descendant, descendant_or_self, ancestor, following_sibling …

### 3.4 Axioms

Set of axioms A of theory db includes the classical axiom schemata of first order logic with equality plus some proper axioms. We divide these proper axioms into the following two sets:

− the set F of atomic facts recorded in the database.
− the set of formulae allowing us to derive facts belonging to the tree geometry predicates.

The sample database we shall use throughout this paper includes the document in figure 4:

$$F = \begin{Bmatrix} \text{node}(/,/), \text{node}(n_1,\text{patients}), \text{node}(n_2,\text{franck}), \text{node}(n_3,\text{service}), \\ \text{node}(n_4,\text{otolarynology}), \text{node}(n_5,\text{diagnosis}), \text{node}(n_6,\text{tonsillitis}), \text{node}(n_7,\text{robert}),... \end{Bmatrix} \quad (1)$$

Axioms allowing us to derive tree geometry facts depend on the numbering scheme. In appendix A, we give the axioms for deriving relations child, descendant, descendant_or_self, preceding_sibling, and immediate_preceding_sibling. These axioms are based on the scheme defined in section 3.2.

The following child relation can be derived from these axioms:

$$\{\text{child}(n_1,/), \text{child}(n_2,n_1), \text{child}(n_3,n_2), \text{child}(n_5,n_2), \text{child}(n_4,n_3), \text{child}(n_6,n_5), \text{child}(n_7,n_1),...\}$$

### 3.5 XPath and XUpdate

We use the following three place xpath predicate to logically interpret XPath expressions:

− xpath(p,n,v), reads "node with label v identified by number n is addressed by path p"

---

[2] More precisely, it should read, "node identified by number x is a child of node identified by number y".

Since semantics of `XPath` is well known, axioms interpreting the `xpath` predicate are given in appendix only (appendix B).

Updating `XML` data is still a research issue (e.g. see [22][20][3]). Today, `XUpdate` is a solution to update `XML` data. The reader may refer to [15] for a complete description of `XUpdate`.

Throughout this section, we shall use the following notations:
- From the logical point of view, whenever we update the database we obtain a new logical theory representing the updated database. Let $db_{new}$ be the new logical theory representing the updated database.
- Let $predicate_{db}$ representing the predicate `predicate` in the theory $db$. Let $predicate_{db_{new}}$ representing the same predicate in the theory $db_{new}$.

For each `XUpdate` operation, we shall give the logical formulae that allow us to derive the theory $db_{new}$ from the theory $db$.

### 3.5.1 Updating node operations

There are two `XUpdate` instructions for updating `XML` nodes: `xupdate:update` and `xupdate:rename`. `xupdate:update` can be used to update the content of existing element nodes. `xupdate:rename` allows attribute or element nodes to be renamed. Both operations need two parameters: the path **PATH** selecting the nodes to update and the new label $v_{NEW}$.

**`xupdate:rename`**: The following two formulae allow us to derive facts belonging to the new set F after an `xupdate:rename` operation.

$$\forall n \forall v, node_{db}(n,v) \wedge \neg xpath_{db}(\mathbf{PATH},n,v) \rightarrow node_{db_{new}}(n,v) \quad (2)$$

Label of nodes which are not addressed by **PATH** are not updated.

$$\forall n \forall v, xpath_{db}(\mathbf{PATH},n,v) \rightarrow node_{db_{new}}(n,\mathbf{V_{NEW}}) \quad (3)$$

Label of nodes which are addressed by **PATH** are updated to $V_{NEW}$.

Example:
Let us consider the operation `xupdate:rename` which *renames* all nodes `service` in `department`:
- **PATH** = //service
- $V_{NEW}$ = department

From formulae 2 and 3, we can derive the new set F:

$$F = \begin{Bmatrix} node(/,/), node(n_1, patients), node(n_2, franck), node(n_3, \mathbf{department}), \\ node(n_4, otolaryngology), node(n_5, diagnosis), node(n_6, tonsillitis), node(n_7, robert), \dots \end{Bmatrix}$$

**`xupdate:update`**: The following two formulae allow us to derive facts belonging to the new set F after an `xupdate:update` operation.

$$\forall n \forall v, node_{db}(n,v) \wedge \neg \exists n' \exists v', (xpath_{db}(\mathbf{PATH},n',v') \wedge child_{db}(n,n')) \rightarrow node_{db_{new}}(n,v) \quad (4)$$

Children of nodes which are not addressed by **PATH** are not updated.

$$\forall n \forall n' \forall v', xpath_{db}(\mathbf{PATH},n',v') \wedge child_{db}(n,n') \rightarrow node_{db_{new}}(n,\mathbf{V_{NEW}}) \quad (5)$$

Children of nodes which are addressed by **PATH** are updated to $V_{NEW}$.

Example:
Let us consider the operation `xupdate:update` which *updates* diagnosis of franck in pharyngitis:
- **PATH** = /patients/franck/diagnosis

- $V_{NEW}$ = pharyngitis

From formulae 2 and 3, we can derive the new set F:

$$F = \begin{Bmatrix} \text{node}(/,/), \text{node}(n_1,\text{patients}), \text{node}(n_2,\text{franck}), \text{node}(n_3,\text{service}), \\ \text{node}(n_4,\text{otolaryngology}), \text{node}(n_5,\text{diagnosis}), \text{node}(n_6,\textbf{pharyngitis}), \text{node}(n_7,\text{robert}),\ldots \end{Bmatrix}$$

### 3.5.2 Creating node operations

There are three XUpdate instructions for creating XML fragments: xupdate:insert-before, xupdate:insert-after and xupdate:append. xupdate:insert-before can be used to insert a new tree as the *immediate* preceding sibling of existing nodes. xupdate:insert-after can be used to insert a new tree as the *immediate* following sibling of existing nodes. xupdate:append can be used to insert a new tree as the last child of existing nodes. All these operations need two parameters: a path **PATH** selecting some nodes and the tree **TREE** to insert. Let us assume **node_TREE** be the two-place predicate used to represent the tree to insert.

$$\forall n \forall v, \text{node}_{db}(n,v) \rightarrow \text{node}_{db_{new}}(n,v) \qquad (6)$$

If a node belongs to the original document then it belongs to the final document. Recall that, thanks to our persistent numbering scheme, identifiers of existing nodes remain the same after an insertion or a deletion.

$$\forall n \forall v \forall n' \forall v' \forall n'' \forall o, \textbf{node}_{\textbf{TREE}}(\textbf{n}',\textbf{v}') \wedge \text{xpath}_{db}(\textbf{PATH},n,v) \wedge \text{create\_number}(n,n',o,n'') \qquad (7)$$
$$\rightarrow \text{node}_{db_{new}}(n'',v')$$

The tree to insert shall be inserted as the last subtree of *each* node selected by **PATH** (append), or as a new preceding-sibling tree of each node selected by **PATH** (insert-before), or as a new following-sibling tree of each node selected by **PATH** (insert-after). Therefore, each node **n'** belonging to the tree to insert is inserted at as many places as nodes addressed by **PATH**. Created numbers n'' assigned to inserted nodes are given by the create_number predicate.

- create_number(n,n',o,n''), reads "node n' is inserted with number n'' by operation o on node n". o can be append, insert-before or insert-after.

Axioms for deriving facts belonging to the create_number predicate depend on the numbering scheme which is used. Axioms implementing the dynamic numbering procedure described in section 3.2.2 are not given in this paper.

Example:
Let us consider the operation xupdate:insert-before which inserts a new medical record:
- **PATH** = /patients/robert
- The tree **TREE** to insert is the following:

$$\{\text{node}(n'_1,\text{albert}), \text{node}(n'_2,\text{service}), \text{node}(n'_3,\text{cardiology}), \text{node}(n'_4,\text{diagnosis}),\}$$

From formulae 6 and 7, we can derive the new set F:

$$F = \begin{Bmatrix} \text{node}(/,/), \text{node}(n_1,\text{patients}), \text{node}(n_2,\text{franck}), \text{node}(n_3,\text{service}), \\ \text{node}(n_4,\text{otolaryngology}), \text{node}(n_5,\text{diagnosis}), \text{node}(n_6,\text{tonsillitis}) \\ \text{node}(n''_1,\text{albert}), \text{node}(n''_2,\text{service}), \text{node}(n''_3,\text{cardiology}), \text{node}(n''_4,\text{diagnosis}), \\ \text{node}(n_7,\text{robert}),\ldots \end{Bmatrix}$$

From the tree geometry axioms we can derive:

$$\begin{cases} \text{preceding\_sibling}(n_1'', n_7), \text{preceding\_sibling}(n_2, n_1''), \text{preceding\_sibling}(n_2'', n_4''), \\ \text{child}(n_1'', n_1), \text{child}(n_2'', n_1''), \text{child}(n_3'', n_2''), \text{child}(n_4'', n_1'') \end{cases}$$

From the dynamic numbering procedure defined in section 3.2.2, we obtain[3]:

$$\{n_1'' = (1,(1,1),(3,2)), n_2'' = (2,(3,2),(7,3)), n_4'' = (2,(3,2),(8,3)), n_3'' = (3,(7,3),(5,2))\}$$

### 3.5.3 Deleting node operations

There is one `XUpdate` instruction for deleting `XML` nodes: `xupdate:remove`. `xupdate:remove` can be used to delete existing subtrees. It requires one parameter: the path **PATH** selecting subtrees to delete.

The following formula allows us to derive facts belonging to the new set F after an `xupdate:remove` operation.

$$\forall n \forall v, \text{node}_{db}(n, v) \land \text{undeleted}(n) \rightarrow \text{node}_{db_{new}}(n, v) \tag{8}$$

If a node belongs to the original document and if it does not belong to a deleted subtree then it belongs to the final document.

− `undeleted(n)`, reads "node n does not belong to a deleted subtree"

We can derive facts belonging to the `undeleted` predicate from the following formulae:

$$\forall n \forall v, \text{node}_{db}(n, v) \land \neg \exists n' \exists v', \begin{cases} \text{descendant\_or\_self}_{db}(n, n') \\ \land \text{xpath}_{db}(\mathbf{PATH}, n', v') \end{cases} \rightarrow \text{undeleted}(n) \tag{9}$$

This formula says that nodes which are not deleted are the nodes which do not belong to a subtree whose root node is addressed by path **PATH**.

Example:

Let us consider the `xupdate:remove` operation which *removes* element diagnosis from franck's medical file:

− **PATH** = `/patients/franck/diagnosis`

From formula 8 and 9, we can derive the new set F:

$$F = \begin{cases} \text{node}(/,/), \text{node}(n_1, \text{patients}), \text{node}(n_2, \text{franck}), \text{node}(n_3, \text{service}), \text{node}(n_4, \text{otolaryngology}), \\ \text{node}(n_7, \text{robert}), \dots \end{cases}$$

## 4. Secure **XML** database

We extend the logical theory `db` to represent a secure `XML` database.

---

[3] The procedure in section 3.2.2 describes the insertion of one single node. For the sake of simplicity, we did not give the rule specifying how identifiers are generated when there is a *simultaneous* insertion of several consecutive nodes at a given level (e.g identifiers $n''_2$ and $n''_4$)

### 4.1 Extended theory

We extend the language with predicates `subject` and `isa` for representing the subject hierarchy,
- `subject(s)`, reads "s is a subject"
- `isa(s,s′)`, reads "subject s is a subject s'"

We also introduce the predicate `rule` for writing the security policy:
- `rule(accept,r,p,s,t)`, reads "subject s is granted privilege r on nodes addressed by path p". t is the priority of the rule.
- `rule(deny,r,p,s,t)`, reads "subject s is denied privilege r on nodes addressed by path p". t is the priority of the rule.

Since accept and deny rules may conflict with each other, we define predicate `perm` to represent the actual privileges held by the subjects:
- `perm(s,r,n)`, reads "subject s is (definitely) granted privilege r on node n"

We extend our theory with the following sets:
- the set `S` of formulae representing the subjects recorded in the database,
- the set $R_S$ of formulae allowing us to derive the subject hierarchy,
- the set `P` of atomic formulae representing the security policy,
- the set $R_P$ of formulae allowing us to solve conflicts between security rules,

### 4.2 Subjects

Let us consider the subjects hierarchy at figure 5. In each tree, internal nodes are *roles* [17] and external nodes are *users*.

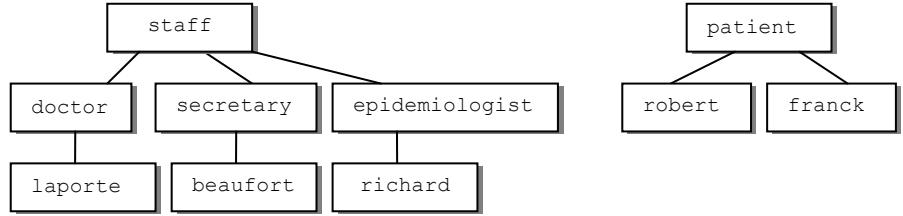

**Fig. 5.** Subject Hierarchy

The following set `S` represents this hierarchy:

$$S = \begin{Bmatrix} \text{subject}(\text{staff}), \text{subject}(\text{secretary}), \text{subject}(\text{doctor}), \text{subject}(\text{epidemiologist}), \\ \text{subject}(\text{patient}), \text{subject}(\text{beaufort}), \text{subject}(\text{laporte}), \text{subject}(\text{richard}), \\ \text{subject}(\text{robert}), \text{subject}(\text{franck}), \text{isa}(\text{secretary},\text{staff}), \text{isa}(\text{doctor},\text{staff}), \\ \text{isa}(\text{epidemiologist},\text{staff}), \text{isa}(\text{laporte},\text{doctor}), \text{isa}(\text{beaufort},\text{secretary}), \\ \text{isa}(\text{richard},\text{epidemiologist}), \text{isa}(\text{robert},\text{patient}), \text{isa}(\text{franck},\text{patient}) \end{Bmatrix} \quad (10)$$

Set $R_S$ includes the two following axioms allowing us to derive the reflexive and transitive closure of the `isa` relation:

$$\forall s, \ \text{subject}(s) \rightarrow \text{isa}(s,s) \quad (11)$$

$$\forall s \forall s' \forall s'', \text{isa}(s,s') \wedge \text{isa}(s',s'') \rightarrow \text{isa}(s,s'') \tag{12}$$

### 4.3 Security Policy

The security policy may refer to the following privileges:
`{position, read, delete, insert, update}`
- if user `s` holds the `position` privilege on node `n` then user `s` has the right to know the existence of `n`.
- if user `s` holds the `read` privilege on node `n` then user `s` has the right to see node `n`.
- if user `s` holds the `insert` privilege on node `n` then user `s` has the right to add a new sub-tree to node `n`.
- if user `s` holds the `update` privilege on node `n` then user `s` has the right to update node `n` (i.e. change its label).
- if user `s` holds the `delete` privilege on node `n` then user `s` has the right to delete the sub-tree of which node `n` is the root.

Privileges should not be confused with *operations*. Operations need privileges to complete. For example, both `xupdate:append` and `xupdate:insert-before` need the `insert` privilege to complete.

Let us now consider the example of security policy defined by axiom `13`. First rule states that staff members have the privilege to `read` the whole document. Second rule partially denies that right from secretaries. Indeed, secretaries are denied the right to see diagnosis. However, rule `3` states that secretaries may know whether the patient was diagnosed or not. Rule `4` and rule `5` state that patients may access their own medical file (`$USER` is a variable containing the session user login name). Rule `6` and rule `7` state that epidemiologists are forbidden to see patient names. Rule `8` states that secretaries may insert new medical files. Rule `9` states that secretaries may update patient names. Rules `10`, rule `11` and rule `12` state that doctors can pose/update/delete a diagnosis.

$$P = \begin{bmatrix} 1.\ \text{rule}(\text{accept},\text{read},//*,\text{staff},10), \\ 2.\ \text{rule}(\text{deny},\text{read},//\text{diagnosis}/*,\text{secretary},11), \\ 3.\ \text{rule}(\text{accept},\text{position},//\text{diagnosis}/*,\text{secretary},12), \\ 4.\ \text{rule}(\text{accept},\text{read},/\text{patients},\text{patient},13), \\ 5.\ \text{rule}(\text{accept},\text{read},/\text{patients}/\text{descendant-or-self}::*[\$USER],\text{patient},14) \\ 6.\ \text{rule}(\text{deny},\text{read},/\text{patients}/*,\text{epidemiologist},15) \\ 7.\ \text{rule}(\text{accept},\text{position},/\text{patients}/*,\text{epidemiologist},16) \\ 8.\ \text{rule}(\text{accept},\text{insert},/\text{patients},\text{secretary},17), \\ 9.\ \text{rule}(\text{accept},\text{update},/\text{patients}/*,\text{secretary},18), \\ 10.\ \text{rule}(\text{accept},\text{insert},//\text{diagnosis},\text{doctor},19), \\ 11.\ \text{rule}(\text{accept},\text{update},//\text{diagnosis}/*,\text{doctor},20), \\ 12.\ \text{rule}(\text{accept},\text{delete},//\text{diagnosis}/*,\text{doctor},21) \end{bmatrix} \tag{13}$$

From practical point of view, we assume that the security administrator inserts these rules one by one. In case of conflict, the last issued rule has the *priority* over the previous ones. The timestamp indicating when the command was issued plays the priority role. For example, rule `1` which says that staff members have the permission to see the whole document is partially cancelled by rule `2` which partially denies that right to some staff members (secretaries).

Set $R_P$ includes the following axiom allowing us to solve the conflicts between the rules and to derive the *actual privileges* held by each subject:

$$\forall s \forall s' \forall r \forall p \forall t \forall n \forall v, \mathrm{isa}(s,s') \wedge \mathrm{rule}(\mathrm{accept},r,p,s',t) \wedge \mathrm{xpath}(p,n,v)$$

$$\wedge \neg \exists s'' \exists p' \exists t', \begin{pmatrix} \mathrm{isa}(s,s'') \wedge \mathrm{rule}(\mathrm{deny},r,p',s'',t') \\ \wedge \mathrm{xpath}(p',n,v) \wedge (t' > t) \end{pmatrix} \rightarrow \mathrm{perm}(s,n,r) \quad (14)$$

This axiom says that if there is an `accept` rule applying to privilege `r`, subject `s` and node `n` and if there is no subsequent `deny` rule applying to privilege `r`, subject `s` and node `n` then subject `s` holds privilege `r` on node `n`.

### 4.4 Access Controls

The purpose of this section is only to illustrate the semantics of the security policy. In section 4.4.1, we state axioms allowing us to derive the view of the source document that users are permitted to see. In section 4.4.2, we state axioms allowing us to derive the new database content after an update.

In this section we do *not* address the problem of how queries should actually be enforced. Regarding this issue, let us, however mention there are basically two approaches:

- The first approach consists of computing and materializing the view of the source document the user is permitted to see [1][7][11]. Queries are then evaluated on the view.
- The second approach [23][30][31] consists of computing the view of the data the user is permitted to see. The user expresses queries by looking at her view. Queries are then rewritten according to the privileges held by the user and evaluated on the source document.

The main drawback of the first approach is that it is expensive to materialize and maintain a large collection of views. The main drawback of the second approach is query rewriting itself which has proved to be a challenging problem since no solution addressing the full XPath language has been proposed yet. In terms of security both approaches should, of course, be equivalent i.e. the same query should return the same answer with both solutions.

Regarding updates, we could easily figure out the two similar approaches:

- The first approach consists of selecting the nodes to update from the view the user is permitted to see. Corresponding source nodes are then retrieved and updated.
- The second approach consists of rewriting queries selecting the nodes to update according to the privileges held by the user. Nodes to update are then selected from the source document and updated.

### 4.4.1 Read Access Controls

The purpose of this section is to define link axioms allowing us to derive the view of the source document that subjects are permitted to see. Each view is represented by a logical theory. Let us denote by `s` the current session user. Let us denote by `view` the theory representing the view that user `s` is permitted to see. The view access control strategy of our model can be informally described as follows:

– If user `s` holds either a `read` or a `position` privilege on node `n` and the parent of node `n` is itself a selected node then node `n` is selected by the view access control mechanism. Axioms `15`, `16` and `17` implement that principle. The fact that a node requires its parent to be selected, in order for it to be selected, shows that the view is a *pruned* version of the source document.

– A selected node for which user `s` holds only the `position` privilege is shown with the `RESTRICTED` label. Axiom `17` implements that principle.

Note that *selected nodes are not renumbered in the view*. This cannot lead to inference channels since numbers are for internal processing only and are not visible to users.

The following formula allows us to derive facts belonging to the view user `s` is permitted to see:

$$\text{node}_{\text{view}}(/,/) \tag{15}$$

This formula says that $\text{node}_{\text{view}}(/,/)$ always belong to the view regardless of user privileges.

$$\forall n \forall v \forall n' \forall v' \forall s, \text{node}_{\text{db}}(n,v) \wedge \text{logged}(s) \wedge \text{perm}(s,n,\text{read}) \wedge \text{child}_{\text{db}}(n,n') \wedge \text{node}_{\text{view}}(n',v') \tag{16}$$
$$\rightarrow \text{node}_{\text{view}}(n,v)$$

This formula says that if the current session user has the permission to *read* node `n` with label `v` and if the parent of node `n` is itself a selected node then the access control mechanism selects node `n` with label `v`.

$$\forall n \forall v \forall n' \forall v' \forall s, \text{node}_{\text{db}}(n,v) \wedge \text{logged}(s) \wedge \text{perm}(s,n,\text{position})$$
$$\wedge \neg \text{perm}(s,n,\text{read}) \wedge \text{child}_{\text{db}}(n,n') \wedge \text{node}_{\text{view}}(n',v') \rightarrow \text{node}_{\text{view}}(n,\text{RESTRICTED}) \tag{17}$$

This formula says that if current session user has the permission to *know the existence* of node `n` and if the parent of node `n` is itself a selected node then the access control mechanism selects node `n` with label `RESTRICTED`. If the session user also holds the `read` privilege then this axiom does not apply.

– `logged(s)`, reads "`s` is the current session user"

Axioms `15`, `16` and `17` can, of course, be implemented by a tree traversal algorithm. In [10], we give such an algorithm.

We can now derive the view of the sample source database (see axiom `1`) each subject is permitted to see.

View for secretaries is the following:

$$\left\{ \begin{array}{l} \text{node}(/,/), \text{node}(n_1,\text{patients}), \text{node}(n_2,\text{franck}), \text{node}(n_3,\text{service}), \text{node}(n_4,\text{otolarynology}), \\ \text{node}(n_5,\text{diagnosis}), \text{node}(n_6,\text{RESTRICTED}), \text{node}(n_7,\text{robert}), \ldots \end{array} \right\}$$

Secretaries can see everything except the content of diagnosis elements. If the diagnosis is posed, then they are provided with the `RESTRICTED` label.

View for patient Robert is the following:

$$\begin{bmatrix} \text{node}(/,/), \text{node}(n_1,\text{patients}), \text{node}(n_7,\text{robert}), \text{node}(n_8,\text{service}), \text{node}(n_9,\text{pneumology}), \\ \text{node}(n_{10},\text{diagnosis}), \text{node}(n_{11},\text{penumonia}) \end{bmatrix}$$

Robert is the current session user. As a patient he has access to its medical file only.

View for epidemiologists is the following:

$$\begin{bmatrix} \text{node}(/,/), \text{node}(n_1,\text{patients}), \text{node}(n_2,\text{RESTRICTED}), \text{node}(n_3,\text{service}), \\ \text{node}(n_4,\text{otolarynology}), \text{node}(n_5,\text{diagnosis}), \text{node}(n_6,\text{tonsillitis}), \text{node}(n_7,\text{RESTRICTED})... \end{bmatrix}$$

Epidemiologists can see everything except patient names.

Doctors can see everything without restriction. Therefore, view for doctors includes the whole database represented by axiom 1.

### 4.4.2 Write Access Controls

The purpose of this section is to define the link axioms allowing us to derive the new database content after an update. The link axioms must take into account the privileges of the user performing the update.

Each `XUpdate` operation requires the path **PATH** parameter to select nodes to update. In order to avoid the vulnerability we described in section 2.5, nodes to update are selected from the view the user is permitted to see. This has the following implications:

- Users cannot perform write operations on nodes they cannot see.
- Since users express `XUpdate` operations by looking at their view, **PATH** parameter might include some node tests equal to `RESTRICTED`.

Note that only the "selecting nodes" step is performed on the view. Thanks to their identifiers, corresponding database nodes are then retrieved and updated.

Let `s` be the current session user submitting the `XUpdate` operation. Let `n` be one of the nodes selected by **PATH**.

- `xupdate:rename`: user `s` needs the update privilege on node `n`.
- `xupdate:update`: user `s` needs the update privilege on the child node of node `n`.
- `xupdate:append`: user `s` needs the insert privilege on node `n`.
- `xupdate:insert-before`: user `s` needs the insert privilege on the parent of node `n`.
- `xupdate:insert-after`: user `s` needs the insert privilege on the parent of node `n`.
- `xupdate:remove`: user `s` needs the delete privilege on node `n`.

Before defining the link axioms allowing us to derive the new database after an update, we need to consider the following issues:

- Each `XUpdate` operation may address several nodes via the **PATH** parameter. Depending on the privileges held by the user submitting the operation, the `XUpdate` operation may succeed for some nodes and fail for others.
- Let us consider an `xupdate:rename` operation addressing a node `n`. Let us assume node `n` is shown in the user's view with a `RESTRICTED` label. Since renaming node `n` would update the original label that the user is not permitted to see, we enforce that nodes which are shown with `RESTRICTED` label cannot be updated.
- Let us consider the `xupdate:update` operation. This operation requires that the user holds the update privilege on the child of each selected node `n`. In fact, the

`xupdate:update` operation on node n is equivalent to the `xupdate:rename` operation on the child of node n. Therefore, the child of node n has to belong to the user's view with its original label, that is, the user needs to hold the `read` privilege on node n.

- Let us consider an `xupdate:remove` operation addressing a node n. If the user removes node n then he actually deletes the subtree of which node n is the root. Some of the nodes which belong to that subtree may not be visible (i.e. may not belong to the user's view). Shall we reject the operation if some nodes of the deleted subtree do not belong to the user's view? On one hand, it would preserve the integrity of data the user is not permitted to see. On the other hand, it would reveal to the user the existence of data she is not permitted to see. In fact there is no definite answer to this question. This is typically a case of conflict between confidentiality and integrity. In this paper, we prefer to emphasize the confidentiality that is, the remove operation is accepted (see axiom 25).

Link axioms allowing us to derive the new database after an update are given below. Note that for each axiom we use the $\text{xpath}_{\text{view}}$ predicate for selecting nodes to update from the view:

**`xupdate:rename`**: We need to adapt axioms 2 and 3 as follows:

$$\forall n \forall v \forall s, \text{node}_{\text{db}}(n,v) \land \neg \begin{Bmatrix} \text{xpath}_{\text{view}}(\textbf{PATH}, n, v) \land \text{logged}(s) \\ \land \text{perm}(s, n, \text{update}) \end{Bmatrix} \rightarrow \text{node}_{\text{db}_{\text{new}}}(n,v) \quad (18)$$

Label of nodes which are not addressed by **PATH** or for which the current session user does not hold the update privilege are not updated.

$$\forall n \forall v \forall s, \text{xpath}_{\text{view}}(\textbf{PATH}, n, v) \land \text{logged}(s) \land \text{perm}(s, n, \text{update}) \rightarrow \text{node}_{\text{db}_{\text{new}}}(n, \textbf{V}_{\text{NEW}}) \quad (19)$$

Label of nodes which are addressed by **PATH** and for which the current session user holds the update privilege are updated to $\textbf{V}_{\text{NEW}}$.

**`xupdate:update`**: We need to adapt axioms 4 and 5 as follows:

$$\forall n \forall v \forall s, \text{node}_{\text{db}}(n,v) \land \neg \exists n' \exists v', \begin{Bmatrix} \text{xpath}_{\text{view}}(\textbf{PATH}, n', v') \land \text{child}_{\text{view}}(n, n') \\ \land \text{logged}(s) \land \text{perm}(s, n, \text{update}) \land \text{perm}(s, n, \text{read}) \end{Bmatrix} \\ \rightarrow \text{node}_{\text{db}_{\text{new}}}(n,v) \quad (20)$$

Label of nodes whose parent is not addressed by **PATH** or for which the current session user does not hold both the update privilege and the read privilege are not updated.

$$\forall n \forall v \forall n' \forall v' \forall s, \text{xpath}_{\text{view}}(\textbf{PATH}, n', v') \land \text{child}_{\text{view}}(n, n') \land \text{logged}(s) \\ \land \text{perm}(s, n, \text{update}) \land \text{perm}(s, n, \text{read}) \rightarrow \text{node}_{\text{db}_{\text{new}}}(n, \textbf{V}_{\text{NEW}}) \quad (21)$$

Label of nodes whose parent is addressed by **PATH** and for which the current session user holds both the update and the read privilege are updated to $\textbf{V}_{\text{NEW}}$.

**`xupdate:append`**: We only need to adapt axiom 7:

$$\forall n \forall v \forall n' \forall v' \forall n'' \forall s, \textbf{node}_{\textbf{TREE}}(\textbf{n}', \textbf{v}') \land \text{xpath}_{\text{view}}(\textbf{PATH}, n, v) \land \text{logged}(s) \land \text{perm}(s, n, \text{insert}) \\ \text{create\_number}(n, n', \text{append}, n'') \rightarrow \text{node}_{\text{db}_{\text{new}}}(n'', v') \quad (22)$$

The tree to insert shall appear as the last subtree of each node selected by **PATH** for which the current session user holds the insert privilege.

**`xupdate:insert-before`**: We need to adapt axiom 7 as follows:

$$\forall n \forall v \forall n' \forall v' \forall n'' \forall f \forall s, \mathbf{node_{TREE}}(n', v') \land \mathrm{xpath_{view}}(\mathbf{PATH}, n, v) \land \mathrm{child_{view}}(n, f)$$
$$\land \mathrm{logged}(s) \land \mathrm{perm}(s, f, \mathrm{insert}) \land \mathrm{create\_number}(n, n', \mathrm{insert\text{-}before}, n'') \quad (23)$$
$$\rightarrow \mathrm{node_{db_{new}}}(n'', v')$$

The tree to insert shall appear as the preceding sibling subtree of each node `n` selected by **PATH** provided the current session user holds the insert privilege on the parent of node `n`.

**`xupdate:insert-after`**: We need to adapt axiom 7 as follows:

$$\forall n \forall v \forall n' \forall v' \forall n'' \forall f \forall s, \mathbf{node_{TREE}}(n', v') \land \mathrm{xpath_{view}}(\mathbf{PATH}, n, v) \land \mathrm{child_{view}}(n, f)$$
$$\land \mathrm{logged}(s) \land \mathrm{perm}(s, f, \mathrm{insert}) \land \mathrm{create\_number}(n, n', \mathrm{insert\text{-}after}, n'') \quad (24)$$
$$\rightarrow \mathrm{node_{db_{new}}}(n'', v')$$

The tree to insert shall appear as the following sibling subtree of each node `n` selected by **PATH** provided the current session user holds the insert privilege on the parent of node `n`.

**`xupdate:remove`**: We only need to adapt axiom 9:

$$\forall n \forall v \forall s, \mathrm{node_{db}}(n, v) \land \neg \exists n' \exists v', \begin{Bmatrix} \mathrm{descendant\_or\_self_{db}}(n, n') \\ \land \mathrm{xpath_{view}}(\mathbf{PATH}, n', v') \\ \land \mathrm{logged}(s) \land \mathrm{perm}(s, n', \mathrm{delete}) \end{Bmatrix} \rightarrow \mathrm{undeleted_{db}}(n, v) \quad (25)$$

This formula says that nodes which are not deleted are the nodes which do not belong to a subtree that the current session user has the permission to delete and whose root is addressed by **PATH**.

## 5 Conclusion

In this paper, we gave the formal definition of a secure XML database. We represented the database content, the query language, the modification language, the subject hierarchy, the security policy and the access controls.

All the logical formulae given in this paper are Horn clauses. Based on these clauses, we wrote a prototype in Prolog simulating a secure XML database. The prototype includes a small database, an XPath and XUpdate interpreter, a sample subject hierarchy, a sample security policy and the access control formulae. It can be downloaded from http://www.univ-pau.fr/~gabillon/xmlsecu. The purpose of this prototype was simply to validate the correctness of the axioms given in this paper. Currently, the prototype uses a simplified version of our numbering scheme [12].

The model in [7] addresses the problem of preserving the validity of the views with respect to the DTD of the original document. Indeed, if a required node in the DTD is not shown in the view the user is permitted to see, then that user can infer about the existence of such a hidden node. Their model does not include the possibility of protecting portions of DTDs. Therefore, they suggest applying a loosening transformation to the DTD so that the view becomes valid with respect to the loosened version of the DTD. Our model does not address this issue which actually belongs to the general problem of inference analysis. A DTD is nothing more than a set of integrity constraints. The security administrator might decide to make a (complex) analysis of possible inference channels which may arise because of inconsistencies between the policy applying to the integrity constraints and the policy addressing the data. Such an analysis is sometimes made for databases which require a high confidentiality level such as multilevel databases [27]. In our model, if we use XML schema instead of DTDs then nothing prevents the owner of the schema from granting to other users the permission to see or partially see the integrity constraints. Indeed an XML schema is an XML document. However, the problem of closing inference channels which may arise because of inconsistencies between the policy addressing the schema and the policy addressing the instances is beyond the scope of this paper and remains word to be done.

Finally, let us mention that a number of works [30][32][33] address the problem of protecting relationships between nodes. The model presented in this paper does not offer this possibility. We are, however, planning to include this feature in a future version of this model.

## Appendix A

With the scheme defined in section 3.2, axioms to derive the `child` and the `preceding_sibling` relations are the followings:

$$\text{child}\big((0,/,(1,1)),/\big)$$

The root node is the child of the document node

$$\forall x \forall y \forall z \forall t \forall v \forall v', \text{node}\big([x,y,z],v\big) \wedge \text{node}\big([x+1,z,t],v'\big) \rightarrow \text{child}\big([x+1,z,t],[x,y,z]\big)$$

If the parent code of a node at level `x+1` is the local code of a node at level `x` then the node at level `x+1` is the child if the node at level `x`.

$$\forall x \forall y \forall z \forall t \forall v \forall v', \text{node}\big([x,y,z],v\big) \wedge \text{node}\big([x,y,t],v'\big) \wedge (t > z)$$
$$\rightarrow \text{preceding\_sibling}\big([x,y,z],[x,y,t]\big)$$

If the local code `t` of a node if higher than the local code `z` of another node sharing the same level and the same parent code, then the node of local code `z` if the preceding sibling of the node of local code `t`.

From these `child` and `preceding_sibling` relations, we can derive all other geometry relations.

$$\forall x \forall y, \text{child}(x,y) \rightarrow \text{descendant}(x,y)$$
$$\forall x \forall z \forall y, \text{child}(x,z) \wedge \text{descendant}(z,y) \rightarrow \text{descendant}(x,y)$$
$$\forall x \forall y, \text{descendant}(x,y) \rightarrow \text{descendant\_or\_self}(x,y)$$
$$\forall x \forall v, \text{node}(x,v) \rightarrow \text{descendant\_or\_self}(x,x)$$
$$\forall x \forall y, \text{preceding\_sibling}(x,y) \wedge \neg \exists z \big(\text{preceding\_sibling}(x,z) \wedge \text{preceding\_sibling}(z,y)\big)$$
$$\rightarrow \text{immediate\_preceding\_sibling}(x,y)$$

We could easily define other tree geometry relations.

## Appendix B

We give the logical interpretation of `XPath` expressions. The reader may refer to [4] for a complete description of `XPath` (version 1.0). Our interpretation is not complete although it covers a very large subset of the `XPath` language. We use the following three place xpath predicate to logically interpret `XPath` expressions:

- xpath(p,n,v), reads "node with label v at position n is addressed by path p"

In the following axioms, we assume that variables p and p' contain any path different from /.

$$\text{xpath}(/,/,/)$$

Path / addresses the document node.

$$\forall v, \text{node}\big((0,/,(1,1)), v\big) \to \text{xpath}\big(/\,*, (0,/,(1,1)), v\big)$$

Path /* addresses the root node.

$$\forall n \forall v, \text{node}(n, v) \to \text{xpath}(//\,*, n, v)$$

Path //* addresses all nodes.

$$\forall v, \text{node}\big((0,/,(1,1)), v\big) \to \text{xpath}\big(/\,v, (0,/,(1,1)), v\big)$$

If the root node has label v then it is addressed by path /v.

$$\forall n \forall v, \text{node}(n, v) \to \text{xpath}(//\,v, n, v)$$

Path //v addresses all nodes of label v.

$$\forall n \forall v \forall n' \forall v' \forall p, \text{node}(n, v) \land \text{child}(n, n') \land \text{xpath}(p, n', v') \to \text{xpath}(p\,/\,v, n, v)$$

Path p/v addresses the nodes of label v which are the children of the nodes addressed by path p.

$$\forall n \forall v \forall n' \forall v' \forall p, \text{node}(n, v) \land \text{child}(n, n') \land \text{xpath}(p, n', v') \to \text{xpath}(p\,/\,*, n, v)$$

Path p/* addresses the nodes which are the children of the nodes addressed by path p.

$$\forall n \forall v \forall n' \forall v' \forall p, \text{node}(n, v) \land \text{descendant}(n, n') \land \text{xpath}(p, n', v') \to \text{xpath}(p\,//\,*, n, v)$$

Path p//* addresses the nodes which are the descendant of the nodes addressed by path p.

$$\forall n \forall v \forall p, \text{xpath}(p, n, v) \to \text{xpath}(p//\text{descendant-or-self::}*, n, v)$$

$$\forall n \forall v \forall p, \text{xpath}(p\,//\,*, n, v) \to \text{xpath}(p//\text{descendant-or-self::}*, n, v)$$

Path p//descendant-or-self::* addresses the nodes which are addressed by path p and the nodes which are addressed by path p//*.

$$\forall n \forall v \forall n' \forall v' \forall p \forall p', \text{xpath}(p, n, v) \land \text{xpath}(p\,/\,p', n', v') \land \text{descendant}(n', n) \to \text{xpath}(p[p'], n, v)$$

Path p[p'] addresses the nodes which are addressed by path p and which have a descendant node addressed by path p/p'.

$$\forall n \forall v \forall p \forall i, \text{xpath}(p, n, v) \land \text{position}(p, n, i) \to \text{xpath}(p[i], n, v)$$

Path p[i] addresses all the nodes which are at position i among the nodes addressed by path p.

- position(p,n,i), reads "node n is at position i among the nodes addressed by path p".

We can derive facts belonging to the position predicate from the following formula:

$$\forall n \forall l \forall i, \text{findall}\left(n', \left(\exists v, \text{xpath}\left(p, n', v\right) \wedge \text{preceding\_sibling}\left(n', n\right)\right), l\right) \wedge \text{length}\left(l, i-1\right)$$
$$\rightarrow \text{position}\left(p, n, i\right)$$

The three-place `findall` predicate is a high-order predicate. However, its semantics is well known since it is an ISO standard `Prolog` predicate. For each n, `findall` creates the list l of all the n′ such that,

$$\left(\exists v, \text{xpath}\left(p, n', v\right) \wedge \text{preceding\_sibling}\left(n', n\right)\right)$$

is true. In other words, `findall` creates the list of all the preceding siblings of n which are addressed by path p. The number of preceding siblings of node n is equal to the length of this list. `length` is also a built-in `Prolog` predicate:

- $\text{length}(l, k)$, reads "the length of list l is k".